\documentclass{aa}

\usepackage[switch]{lineno}
\usepackage{graphicx}
\usepackage{txfonts}
\usepackage{multirow}
\usepackage[breaklinks=true]{hyperref}
\usepackage{xcolor}

\begin{document}

\title{Search for gamma-ray variability around \emph{Fermi}-LAT pulsar glitches}
\titlerunning{Gamma-ray variability in Fermi-LAT pulsar glitches}

\author{
G.~Cozzolongo$^{(1)}$ \and 
A.~Fiori$^{(2)}$ \and 
M.~Razzano$^{(2,3)}$ \and 
P.~M.~Saz~Parkinson$^{(4)}$
}

\institute{
\inst{1}~Friedrich-Alexander Universit\"at Erlangen-N\"urnberg, Erlangen Centre for Astroparticle Physics, Nikolaus-Fiebiger-Str. 2, 91058 Erlangen, Germany\\ 
\inst{2}~Istituto Nazionale di Fisica Nucleare, Sezione di Pisa, I-56127 Pisa, Italy\\ 
\inst{3}~Universit\`a di Pisa, Dipartimento di Fisica E. Fermi, I-56127 Pisa, Italy\\ 
\inst{4}~Santa Cruz Institute for Particle Physics, Department of Physics and Department of Astronomy and Astrophysics, University of California at Santa Cruz, Santa Cruz, CA 95064, USA\\ 
\email{giovanni.cozzolongo@fau.de} \\
\email{alessio.fiori@pi.infn.it} \\
\email{massimiliano.razzano@unipi.it} \\
}

\date{Received XXX; accepted XXX}
 
\abstract
{
Pulsars are the most numerous class of Galactic $\gamma$-ray sources detected by the \emph{Fermi} Large Area Telescope (LAT). 
Young pulsars occasionally experience sudden timing discontinuities called glitches, which are characterized by rapid changes in their rotational parameters, and are usually followed by a return to their regular rotation behavior. PSR J2021+4026 is a unique Fermi-LAT pulsar in that it exhibits peculiar, quasiperiodic switches between two states of varying spin-down rates, approximately every 3--4 years. The mode changes in PSR J2021+4026 are correlated with sudden changes in the $\gamma$-ray emission features of the pulsar.
}
{The goal of this study was to search for variability in the emission features of $\gamma$-ray pulsars correlated with the occurrence of glitches. 
We introduce a novel approach to analyzing LAT $\gamma$-ray pulsars, focusing on a systematic examination of variability associated with changes in pulsar spin-down rates.}
{We tracked the changes in the rotation and $\gamma$-ray emission for a selection of seven glitches that we deemed promising based on the observed changes in the spin-down rates. We conducted a binned likelihood spectral analysis using 14 years of \emph{Fermi}-LAT data. We specifically analyzed windows of data collected around the epoch of the selected glitches. We improved the precision on the best-fit parameters by incorporating likelihood weights, which we calculated based on a model of the diffuse background, thereby accounting for systematic error contributions.}
{The phase-averaged flux and spectral parameters of the pulsars analyzed in this study do not vary significantly across any of the glitches investigated. 
The $95\%$ upper limits on the relative flux change from this work indicate that the flux of the Vela pulsar is unchanging, with a 0.5\% upper limit on the relative change, making it a promising candidate for further searches.
The connection between glitch dynamics and $\gamma$-ray emission in pulsars remains unclear, and PSR J2021+4026 remains unique in terms of its gamma-ray variability properties.
We conclude that a comprehensive investigation of glitches with the goal of further unraveling their underlying mechanisms is warranted.
}
{}


\maketitle

\section{Introduction}

Pulsars are highly magnetized and rapidly rotating neutron stars that emit regular pulses of radiation at their spin rate. While the majority of known pulsars have been observed only at radio frequencies, the number of pulsars detected at $\gamma$-ray energies has risen significantly in recent years. 
NASA's \emph{Fermi} Large Area Telescope (LAT; \citealt{Atwood_2009}) set off a revolution in pulsar science.
Since its launch, in June 2008, \emph{Fermi}-LAT has detected almost 300 $\gamma$-ray pulsars~\footnote{\url{https://confluence.slac.stanford.edu/display/GLAMCOG/Public+List+of+LAT-Detected+Gamma-Ray+Pulsars}}, approximately half of which are millisecond pulsars. The Third \emph{Fermi}-LAT Catalog of Gamma-ray Pulsars (3PC; \citealt{2023ApJ...958..191S}) presents confirmed $\gamma$-ray pulsations from 294 pulsars, with an additional 45 pulsars coincident with LAT $\gamma$-ray sources that will likely yield pulsations once phase-connected rotation ephemerides become available.

Some pulsars are known to undergo small sudden increases in their rotation rates, or glitches, followed by a gradual return to their pre-glitch rotation rates. Although much more commonly observed in young and energetic pulsars, such as Vela and the Crab, where they were first detected decades ago \citep{BAYM1969, Boynton69}, there have been reports of glitches being detected in some millisecond pulsars \citep{Cognard04,McKee16}. A number of models have been proposed to explain glitches, including starquakes affecting the surface of the neutron star \citep{Ruderman.1969} and the decoupling of the internal superfluid from the crust \citep{Anderson&Itoh.1975}, among many others \citep{Haskell15, 2022RPPh...85l6901A}.

Variability in the emission features of certain radio pulsars, that is, pulse shape and intensity, has been found to be correlated with the rotational behavior of such pulsars. The rotation observed in these pulsars is often affected by changes in spin-down rates, with some pulsars abruptly switching between states during these episodes \citep{2010Sci...329..408L}. Although observed in the radio, one of the radio pulsars experiencing this switching behavior, PSR J0742-2822, has a LAT $\gamma$-ray counterpart. The precise mechanism responsible for the correlated variability observed during such state-switching episodes remains elusive. In other cases, a correlation between the frequency derivative of a pulsar and changes in the pulse profile shape has been uncovered, using a Gaussian process regression to model noisy observations \citep{Brook16, Lower2025}. Finally, in at least one instance, a potential connection between glitches and changes in the magnetospheric emission has been uncovered \citep{Weltevrede11}. 

In the $\gamma$-ray regime, mode changes have been detected in the LAT $\gamma$-ray pulsar J2021+4026 \citep{2013ApJ...777L...2A, Ng_2016, Takata_2020}. PSR J2021+4026 (hereafter J2021) is a young, radio-quiet $\gamma$-ray pulsar ($P \sim 256\,\mathrm{ms}$) located within the shell of the Gamma Cygni (G 78.2$+$2.1) supernova remnant. The pulsar was discovered with \emph{Fermi}-LAT using blind periodicity searches \citep{2009Sci...325..840A} and is associated with the bright unidentified source EG J2020+4017 observed by the Energetic Gamma Ray Experiment Telescope (EGRET). A likely X-ray counterpart was inferred using \emph{Chandra} data \citep{2011ApJ...743...74W} and X-ray pulsations were subsequently detected with \emph{XMM-Newton} \citep{2013ApJ...770L...9L}. 
J2021 undergoes repeated and quasiperiodic mode changes that affect its period derivative at intervals of 3--4 years. 
Although similar events have been observed in other pulsars, J2021 is unique in displaying rapid $\gamma$-ray flux variability that is simultaneous with its changing spin-down rate. The first state change (jump) was observed in October 2011 \citep{2013ApJ...777L...2A}, when the flux ($F_\gamma\sim7.9\times10^{-10}\,\mathrm{erg\,cm^{-2}\,s^{-1}}$) dropped by 18\% and the spin-down rate ($\Dot{P} \sim 5.4\times10^{-14}\,\mathrm{s\,s^{-1}}$) increased by $5.6\%$ on a timescale of $<7$ days; followed by a slow recovery, over a period of about $100$ days.

Out of the 294 $\gamma$-ray pulsars listed in 3PC, 54 have experienced at least one glitch, with a total of 128 glitches detected in the $\gamma$-ray band. Meanwhile, the Australia Telescope National Facility (ATNF) Pulsar Catalogue \citep{ATNF}\footnote{\url{https://www.atnf.csiro.au/research/pulsar/psrcat/}} 
lists 3268 pulsars with emission in the radio band, of which $194$ are known to have glitched, for a total of 572 glitches (58 from radio-quiet pulsars). This count does not include the $121$ spin-powered pulsars with pulsed emission only at infrared or higher frequencies, $17$ of which are glitching pulsars with $58$ total glitches. Only 29 glitches appear in both the 3PC and ATNF catalogs. Moreover, $8$ events from 3PC are actually spin-down glitches, meaning that they have $\Delta\nu/\nu<0$ at the glitch epoch.

We performed a maximum likelihood spectral analysis of the data encompassing a collection of potential glitches identified by LAT. Our aim is to search for possible variations in the $\gamma$-ray flux and spectral parameters of these pulsars that are associated with simultaneous changes in their pulsar spin-down rates. The manuscript is outlined as follows. In Sect. \ref{sec:Glitch_Sample} we select a sample of glitches in LAT pulsars. In Sec. \ref{sec:Analysis overview 2} we report on the LAT spectral analysis and present the results. We discuss the results in Sect. \ref{sec:Discussion}, which is followed by conclusions in Sect.~\ref{sec:Spectral variability analysis}.

\section{Glitch sample selection criteria}
\label{sec:Glitch_Sample}

Among the timing models associated with 3PC\footnote{Available at \url{https://fermi.gsfc.nasa.gov/ssc/data/access/lat/3rd_PSR_catalog/3PC_HTML/}}, there are 128 instances of {\tt GLEP}, a parameter that indicates the epoch of a putative glitch. Although we took this number as a proxy for the number of glitches in the $\gamma$-ray pulsar population in 3PC, we caution that these do not necessarily represent all the glitches in 3PC pulsars over the time period in question, nor do all of these instances necessarily correspond to true glitches.

To describe the pulsar rotation, the spin frequency is usually expressed by a Taylor series expansion:
\begin{linenomath}
\begin{equation}
    \nu(t)=\nu_0+\Dot{\nu}_0(t-t_0)+\dfrac{1}{2}\Ddot{\nu}_0(t-t_0)^2+\dots,
\end{equation}
\end{linenomath}
where $\nu_0=\nu(t_0)$ is the spin frequency at a reference time (epoch), $t_0$, while $\Dot{\nu}_0=\Dot{\nu}(t_0)$ and $\Ddot{\nu}_0=\Ddot{\nu}(t_0)$ are its time derivatives at the same reference epoch. The observed parameters $\nu$, $\Dot{\nu}$ and $\ddot{\nu}$ are related to the emission processes causing pulsars to spin down. The value of $\Ddot{\nu}$ is usually too small to be measured, except for very young pulsars \citep{2004hpa..book.....L}. Terms of higher order are typically attributed to the effect of pulsar rotational (or timing) noise \citep{Cordes&Helfand.1980, Shannon.2016}. Thus, in the most general case, while undergoing the slow spin-down process, the rotational frequency is given by:
\begin{linenomath}
\begin{equation}
    \nu_{\mathrm{sd}}(t)\equiv\nu_0 + \sum_i\dfrac{\nu_i\,t^i}{i!},
\end{equation}
\end{linenomath}
where we arbitrarily set the glitch epoch at $t = 0$ and sd stands for spin-down.
The post-glitch frequency is written as the sum of the pre-glitch frequency plus the permanent change at the time of the glitch and decay terms:
\begin{linenomath}
\begin{equation}
\nu(t)=
     \begin{cases}
       \nu_{\mathrm{sd}}(t)&t\leq0\\
       \nu_{\mathrm{sd}}(t) + \Delta \nu_\mathrm{p} + \Delta\dot{\nu}_\mathrm{p}\,t+ \dfrac{1}{2}\Delta\ddot{\nu}_p\,t^2 + \Delta\nu_\mathrm{d}\,e^{-\tfrac{t}{\tau_\mathrm{d}}}&t>0,
     \end{cases}
\end{equation}
\end{linenomath}
where $\nu$ is the pulse frequency, $\nu_0$ is its value at the glitch epoch extrapolated from pre-glitch data, $\Delta\nu_p$ and $\Delta\dot{\nu}_p$ are the permanent changes in $\nu$ and $\dot{\nu}$ at the time of the glitch, $\Delta\nu_d$ is the decaying part of the frequency increment at the time of the glitch, and $\tau_d$ is the decay timescale. For $t<0$, $\Delta\nu_p$, $\Delta\dot{\nu}_p$ and $\Delta\nu_d$ are all zero. Note that in general the observed jumps can include contributions from terms with different decay timescales.

\begin{figure}
    \centering
    \includegraphics[width=0.5\textwidth]{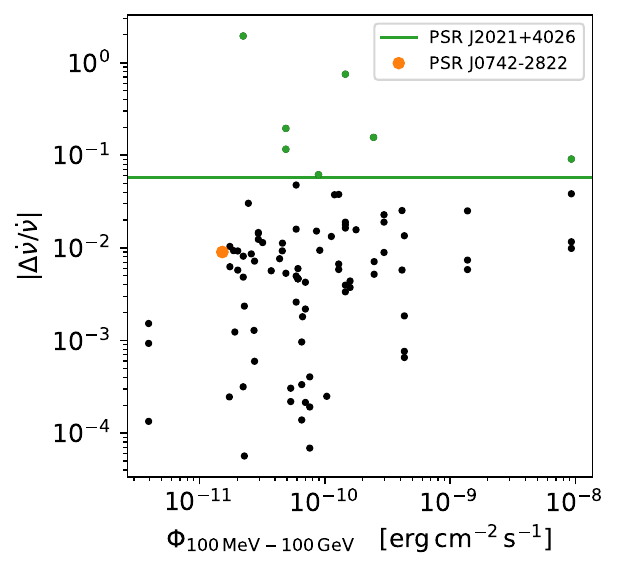}
    \caption{Distribution of $|\Delta\Dot{\nu}/\Dot{\nu}|$ with respect to the 0.1--100 GeV energy flux, according to the 4FGL catalog \citep{2020ApJS..247...33A}. The green line indicates the value for J2021. Green dots indicate glitches with a larger jump than J2021; the orange dot corresponds to PSR J0742-2822, the only glitch with a nonzero jump in spin-down.}
    \label{fig:dataset}
\end{figure}

We selected a subset of glitches that we considered particularly promising for this analysis (see Figure \ref{fig:dataset}). As a sampling criterion, we defined the fractional change in spin frequency as
\begin{linenomath}
\begin{equation}
    \dfrac{\Delta\nu}{\nu} = \dfrac{\nu(0_+)-\nu(0_-)}{\nu(0_-)} = \dfrac{\Delta\nu_p+\Delta\nu_d}{\nu_0} 
\end{equation}
\end{linenomath}
and the fractional change in spin-down rate as
\begin{linenomath}
\begin{equation}
    \dfrac{\Delta\dot{\nu}}{\dot{\nu}} = \dfrac{\dot{\nu}(0_+)-\dot{\nu}(0_-)}{\dot{\nu}(0_-)} = \dfrac{\Delta\dot{\nu}_p-\dfrac{\Delta\nu_d}{\tau_d}}{\dot{\nu}_0},
    \label{fractional-spindown}
\end{equation}
\end{linenomath}
where $\dot{\nu}_0 = \dot{\nu} (t = 0)$. We computed these quantities for each glitching pulsar in 3PC based on their best-fit values. 
We made the assumption that gamma-ray variability is related to the global change in the spin-down rate, that is, to $\Delta\dot{\nu} / \dot{\nu}$. 
Our reference is J2021, 
which shows a permanent change in the spin-down rate $\Delta \dot{\nu}_p / \dot{\nu}_0  \sim 5.8\%$ \citep{2024A&A...685A..70F}.
Therefore, we selected all glitches reported in 3PC with a total $\Delta\Dot{\nu}/\Dot{\nu} > 5.8\%$ 
occurring within the time span of the LAT mission. 
In particular, we excluded the glitches of PSR J1801-2451 at MJD 49469, 52273, and 54470, and one glitch of PSR J2111+4606 at MJD 54750.

Although discarded by our selection criteria, we also included two glitches from PSR J0742-2822, the only switching radio pulsar with a \emph{Fermi}-LAT counterpart \citep{2010Sci...329..408L}.
In Fig. \ref{fig:dataset} we report the values of $\Delta\Dot{\nu}/\Dot{\nu}$ versus the integrated energy flux in the 0.1--100 GeV range.

Table \ref{tab:dataset} shows our sample of selected glitches. For each glitch, we chose two time intervals, in order to test for changes in the observed spectral energy distribution (SED). The two time intervals are defined as follows: the pre-glitch time interval starts at the epoch of a previous glitch, when it exists, or on October 28, 2008 (MJD 54767)\footnote{
For the analysis of SOURCE class photons at energies $>30$ GeV, it is recommended to exclude data prior to 
MJD 54767, as stated in the analysis caveats with LAT Pass 8 data (\url{https://fermi.gsfc.nasa.gov/ssc/data/analysis/LAT_caveats.html})}, and ends at the glitch epoch; the post-glitch time interval starts at the glitch epoch and ends at the epoch of a subsequent glitch, when it exists, 
or at the end date of the timing model validity window. This results in typical time windows spanning between months and years.

The reason for choosing large time windows is linked to the implicitly low count rate of \emph{Fermi}-LAT data. The typical rate of photons observed from a LAT pulsar is up to tens of counts per day over the energy range 100 MeV -- 300 GeV \citep{Abdollahi_2022}. This implies that the statistical significance of the measure of the flux, and thus of any flux change, improves with the observing time. For the same reason, it is not possible to achieve precise flux measures with observing windows comparable to the timescale of glitches. Therefore, we could not localize the epoch of any rapid flux change precisely by means of spectral analysis alone. Consequently, we had to rely on the knowledge of glitch epochs provided by independent timing methods. The analysis of J2021 carried out by \citet{2024A&A...685A..70F} also used time windows of up to $\sim 3.5$ years.

\begin{table*}
  \caption{Glitches selected for the variability analysis.}
     \label{tab:dataset}
     
     \begin{tabular}{cccccc|ccc}
        \hline
        PSR [J2000] & $\nu\,$[Hz] & $\Dot{\nu}\,[-10^{-11}]$ & $\tau_\mathrm{c}\,$[kyr] & $B_\mathrm{S}\,[10^{12}\,\mathrm{G}]$ & $\Dot{E}_\mathrm{rot}\,[-10^{37}\,\mathrm{erg\,s}^{-1}]$ & GLEP$\,$[MJD] & $\Delta\nu/\nu\,[10^{-9}]$ & $\Delta\Dot{\nu}/\Dot{\nu}\,[10^{-3}]$\tabularnewline
        \hline
        \hline
        J0835-4510$^a$ & $11.19$ & $1.55$ & $11.3$ & $3.38$ & $0.69$ & $56556$ & $3091.21$ & $91.09$ \tabularnewline
        J1028-5819 & $10.94$ & $0.19$ & $89.87$ & $1.23$ & $0.083$ & $57853$ & $2342.98$ & $156.18$ \tabularnewline
        J1341-6220 & $5.17$ & $0.68$ & $12.1$ & $7.08$ & $0.14$ & $55042$ & $-724.87$ & $-1942.17$ \tabularnewline
        J1833-1034 & $16.15$ & $5.27$ & $4.85$ & $3.58$ & $3.4$ & $55156$ & $-48.00$ & $-61.74$ \tabularnewline
        J2111+4606 & $6.34$ & $0.65$ & $17.5$ & $4.81$ & $0.14$ & $55668$ & $1376.77$ & $195.1$\tabularnewline
        \hline
        J0742-2822 & $6.00$ & $0.06$ & $158.81$ & $1.69$ & $0.014$ & $55020$ & $103.72$ & $9.02$ \tabularnewline
        & & & & & & $56727$ & $2.94$ & $0.0$ \tabularnewline
        \hline
     \end{tabular}
     
     \tablefoot{
     PSR is the pulsar name in J2000 epoch, $\nu$ is the rotational frequency, $\Dot{\nu}$ is its first time derivative, $\tau_\text{c}$ is the characteristic age, $B_\text{S}$ is the magnetic field intensity at the neutron star surface, $\Dot{E}_\text{rot}$ is the spin-down luminosity, GLEP is the glitch epoch (in MJD), $\Delta\nu/\nu$ is the spin-down jump and $\Delta\Dot{\nu}/\Dot{\nu}$ the spin-down rate jump. $^a$ Vela pulsar.
     }
\end{table*}

\section{Search for $\gamma$-ray variability}
\label{sec:Analysis overview 2}
To search for spectral variations in our sample of $\gamma$-ray pulsars, we performed a maximum likelihood fit to LAT data of Pass 8 SOURCE class photons \citep{2013arXiv1303.3514A, 2018arXiv181011394B}. We included the necessary corrections due to energy dispersion, thus enhancing the accuracy at energies below 1 GeV. We also included likelihood weights, which take into account the systematic errors in the diffuse background into account. Likelihood weights were calculated for each event type, a method known as summed likelihood. Our approach is equivalent to the one adopted by \citet{2024A&A...685A..70F} for the analysis of the mode changes of J2021.
We carried out our analysis using the standard Fermitools~\footnote{\url{https://fermi.gsfc.nasa.gov/ssc/data/analysis/software/}} (formerly ScienceTools), version 1.2.23, released on February 2, 2020. 

\subsection{Data selection} 
\label{subsec:Data selection 2}

\emph{Fermi}-LAT data are publicly available via the Fermi Science Support Center (FSSC)\footnote{\url{https://fermi.gsfc.nasa.gov/ssc/data/access/}}, which enables events to be selected within a defined radius around a chosen position in the sky and within a given time and energy range. 
We selected SOURCE class photons (\texttt{evclass}=$128$) in the energy range $100\,\mathrm{MeV}$ to $300\,\mathrm{GeV}$ at incident zenith angles $<90^\circ$. The corresponding name of the instrument response functions (IRFs) we used is P8R3\_SOURCE\_V2. Due to the exponential shape of the pulsar SED, we expected a cutoff at $\approx1\,\mathrm{GeV}$, with very few counts expected at very high energies.

Because of the large point spread function (PSF) at low energies (about $3.5^\circ$ at $100\,\mathrm{MeV}$), the flux of the analysis target is spread over a wide field and is affected by photons from neighboring sources. Thus, we had to define a region of interest (ROI) that is larger than the PSF in order to model all neighboring sources in addition to the source of interest.
We chose a square ROI of size $15^\circ \times 15^\circ$.
We applied a binning to photon energies and incidence directions to run a binned likelihood analysis.
In the pre-glitch window of the PSR J1341-6220 glitch at MJD 55042 and in both windows the PSR J0742-2822 glitch at MJD 55020, we adopted a bin size of $0.2^\circ$. In all other windows, we used pixels of size $0.1^\circ$. Typically, the FSSC recommends using 8--10 logarithmically spaced energy bins per decade\footnote{\url{https://fermi.gsfc.nasa.gov/ssc/data/analysis/documentation/Pass8_edisp_usage.html}}. Since \emph{Fermi}-LAT point sources, and especially $\gamma$-ray pulsars, are generally very bright up to $10\,\mathrm{GeV}$, we chose ten energy bins per decade, enabling us to better track the fast variations in the PSF and effective area at low energies. This choice resulted in a total of 35 logarithmically spaced bins in the energy range from 0.1--300 $\mathrm{GeV}$.

Since IRFs are functions of the angle between the direction of the source and the LAT normal, the predicted number of counts depends on the amount of time spent at a given inclination angle during the observation. We note that calculating the array of livetimes at all points of the sky is computationally expensive and results in very large files. Because the exposure varies slowly over the whole sky, we chose a pixel size larger than the angular binning of the data, which should not significantly affect our results. However, the angular resolution should accurately scan all inclination angles up to $90^\circ$. Thus, we used $1^\circ$ pixels and $\Delta\cos\theta=0.025$.

\subsection{Flux and spectral monitoring}
\label{sec:Flux and timing monitoring}
Our \emph{Fermi}-LAT analysis was carried out using the \texttt{pyLikelihood}\footnote{\url{https://fermi.gsfc.nasa.gov/ssc/data/analysis/scitools/python_usage_notes.html}} package, a Python implementation of Fermitools. The best fits were obtained using the NEWMINUIT \citep{JAMES1975343} optimizer.
To improve the accuracy of the output from the likelihood analysis, we used a summed likelihood method, which consisted of preparing different sets of data and including them in the analysis as separate components. Each component was treated independently, according to their respective IRFs. In this way, lower quality data will have a lower impact on the ultimate outcome of the fit. In our analysis, we used four components, one for each PSF event type. Whenever the statistical precision on the Galactic diffuse emission was better than the systematic variations in fluxes in the ROI, we chose to neglect the faintest features in the ROI. Thus, we implemented our likelihood weights with the purpose of obtaining test statistic (TS) values that are not overestimated. This is done by properly handling the systematics on the diffuse emission model \citep{2020ApJS..247...33A}. Weighted likelihood was first introduced by \citet{10.1214/009053604000001309}. The overall effect is an increase in the uncertainties, since the precision is now limited by the systematics. The target systematic error, $\epsilon$, is typically about $3\%$.

For each target, we built a model of the ROI starting from the LAT 14-year Source Catalog, 4FGL-DR4 \citep{2023arXiv230712546B}. Due to the large LAT PSF, in order to account for the flux contribution from sources outside the ROI, we included all 4FGL sources within $20^\circ$ of the pulsar.
We modeled the $\gamma$-ray spectrum of our pulsars using the \texttt{PLSuperExpCutoff4}\footnote{\url{https://fermi.gsfc.nasa.gov/ssc/data/analysis/scitools/source_models.html}}
(PLEC4) function, which is a power law with an exponential cutoff,
\begin{linenomath}
\begin{equation}
    \dfrac{dN}{dE}=N_0\biggl(\dfrac{E}{E_0}\biggl)^{d/b-\Gamma_\mathrm{S}}\exp\biggl\{\dfrac{d}{b^2}\biggl[1-\biggl(\dfrac{E}{E_0}\biggl)^b\biggr]\biggr\}
\end{equation}
\end{linenomath}
where the normalization $N_0$ is directly the flux density at the reference energy $E_0$,
and the shape parameters are the spectral slope, $\Gamma_\mathrm{S} = d\log(dN/dE)/d\log E$, and the spectral curvature, $d = d^2\log(dN/dE)/d(\log E)^2$, at $E_0$. The case in which the exponential index $b$ is zero corresponds to a particular case of the model \texttt{LogParabola}. This parameterization, devised for the LAT 12-year Source Catalog, 
4FGL-DR3 \citep{Abdollahi_2022}, is more stable and reduces the correlation between parameters, compared to the models used in previous catalogs.
We included the background models\footnote{\url{https://fermi.gsfc.nasa.gov/ssc/data/access/lat/BackgroundModels.html}} for the Galactic diffuse emission 
and isotropic diffuse emission for each event type. 

For each glitch in Table~\ref{tab:dataset}, we set the free parameters based on the following criteria.
Since we needed to model the brightest sources in the ROI, the normalizations of the main extended sources in the field and the bright pulsars in the ROI were kept free. 
The \emph{Fermi}-LAT PSF is about $3.5^\circ$ at $100\,\mathrm{MeV}$; thus, sources within this distance of our target may contribute low energy counts. In order to model them, we freed the normalization of all sources within $3.5^\circ$ of the center of the ROI.
We also had to account for variable sources, whose fluxes can vary significantly from the average value reported in the catalog.  
In 4FGL, sources with a variability index $\rm{TS}_{\rm{var}} > 18.48$ are considered variable over one-year timescales with $99\%$ confidence \citep{2020ApJS..247...33A}. 
Thus, we freed the normalizations of all sources with $\rm{TS}_{\rm{var}}$ above this value.
However, the efficiency and the speed of the optimization algorithm critically depend on the number of free parameters. To avoid adding too many degrees of freedom to our fit, we fixed all sources with $\mathrm{TS}<25$, which corresponds to a detection significance lower than $4\,\sigma$ \citep{2020ApJS..247...33A}.
We multiply multiplied the Galactic diffuse emission by a power law so that the SED could adjust to absorb part of the large-scale residuals. The normalization and index of the power law were kept free. The isotropic emission was kept fixed.
The parameters $d$ and $b$ were fixed. With these criteria, the amount of free parameters is typically between 20 and 30.

The analysis of some of the selected pulsars required some tuning due to the intrinsic complexity of their regions. In particular, the region around Vela showed significant negative residuals, an indication that the model for this source \citep{FermiVela, FermiVela-X, 2020ApJS..247...33A} poorly represents the $\gamma$-ray emission.
We do not find evidence of flux variability in any of the selected targets. We computed the 95\% confidence level upper limits on the flux variation, both absolute and relative to the pre-glitch flux. The results are shown in Table \ref{tab:results}.

\section{Discussion} \label{sec:Discussion}
We conducted the first search for variability in $\gamma$-ray emission associated with pulsar glitches characterized by the highest spin-down rates. Our approach aimed to uncover novel insights into the relationship between glitch dynamics and electromagnetic emission in the $\gamma$-ray spectrum. This pursuit was motivated by the observation that the first variable pulsar detected in the $\gamma$-ray spectrum exhibited a notable change in its spin-down rate. Although this change was not strictly categorized as a glitch, glitches typically entail significant shifts in spin-down rates, making them targets for exploring such variations. We have presented an analysis of seven glitches in six \emph{Fermi}-LAT pulsars.
We estimated the upper limits on flux variations by calculating the absolute difference between the pre-glitch and post-glitch fluxes. The 0.95 quantile of this distribution of absolute differences was taken as the 95\% upper limit on the flux variation.
Our analysis yielded 95\% upper limits on the absolute flux variations associated with these glitches.

We discuss two peculiar pulsars among our chosen targets.
PSR J0835-4510, better known as the Vela Pulsar, is the brightest $\gamma$-ray pulsar \citep{2022ApJ...934...30K}, one of the brightest radio pulsars, and is also detected in the optical and X-ray band. It is located nearby, at a distance of $287\,\mathrm{pc}$. It has a spin-down power of $6.9\times10^{36}\,\mathrm{erg\,s^{-1}}$ 
and a surface magnetic field of $3.38\times10^{12}\,\mathrm{G}$.
The Vela pulsar is associated with the Vela supernova remnant ($\geq10\,\mathrm{kyr}$ old; \citealt{Sushch_2014}) and has been extensively monitored since its discovery in 1968 \citep{1968Natur.220..340L}. The Vela pulsar shows the lowest relative upper limit, suggesting that, if any flux variation occurred, it would be a good candidate for a detection. This is consistent with Vela being one of the brightest and most closely studied $\gamma$-ray pulsars.

The second peculiar target is PSR J0742-2822, which is known to exhibit quasiperiodic changes in its observed pulse profile and spin-down rate \citep{Dang_2021}. It is a radio-loud pulsar with rotational frequency of 6 $\mathrm{Hz}$, a spin-down rate of $-0.06\times10^{-11}\,\mathrm{Hz\,s^{-1}}$, a characteristic age of $158.81\,\mathrm{kyr}$, a spin-down power of $1.4\times 10^{35}\,\mathrm{erg\,s^{-1}}$, and a surface magnetic field of $1.69\times10^{12}\,\mathrm{G}$. We considered two glitches from this pulsar. \citet{10.1093/mnras/stt660} found no correlation between the observed pulse shape and the spin-down rate for at least 200 days prior to one of these glitches (the one at epoch 55020 MJD), following which the correlation became strong. These observations indicate that changes in emission state can be caused by the interaction between the interior of the neutron star and the magnetosphere of the pulsar.

\begin{table*}
\caption{Results of the variability analysis.}
\label{tab:results}
\scalebox{0.93}{
\begin{tabular}{cccccccc} 
\hline           
PSR [J2000] & GLEP [MJD] & window & time [d] & $\Gamma_\text{S}$ & $\Phi\,[10^{-11}\,\text{erg}\,\text{cm}^{-2}\,\text{s}^{-1}]$ & $\vert\Delta\Phi\vert_{95\%}\,[10^{-11}\,\text{erg}\,\text{cm}^{-2}\,\text{s}^{-1}]$ & $\vert\Delta\Phi\vert_{95\%}/\Phi$ \\
\hline
\hline
\multirow{2}{*}{J0835-4510} & \multirow{2}{*}{56556} & pre & 1147 & $2.262 \pm 0.003$ & $923 \pm 2$ & \multirow{2}{*}{5} & \multirow{2}{*}{0.005}\\
 & & post & 1179 & $2.258 \pm 0.003$ & $915.9 \pm 1.6$ & \\
\hline
\multirow{2}{*}{J1028-5819} & \multirow{2}{*}{57853} & pre & 3086 & $2.367  \pm 0.014$ & $25.3 \pm 0.3$ & \multirow{2}{*}{0.9} & \multirow{2}{*}{0.04}\\
 & & post & 755 & $2.349 \pm 0.016$ & $24.5 \pm 0.3$ & \\
\hline
\multirow{2}{*}{J1341-6220} & \multirow{2}{*}{55042} & pre & 161 & $0.3 \pm 0.6$ & $1.0 \pm 0.8$ & \multirow{2}{*}{2} & \multirow{2}{*}{$>1$}\\
 & & post & 793 & $2.4 \pm 1.8$ & $1.5 \pm 0.3$ & \\
\hline
\multirow{2}{*}{J1833-1034} & \multirow{2}{*}{55156} & pre & 389 & $2.41 \pm 0.07$ & $8.7 \pm 0.7$ & \multirow{2}{*}{2} & \multirow{2}{*}{0.2}\\
 & & post & 3544 & $2.34 \pm 0.02$ & $8.5 \pm 0.6$ & \\
\hline
\multirow{2}{*}{J2111+4606} & \multirow{2}{*}{55668} & pre & 901 & $1.92 \pm 0.05$ & $4.8 \pm 0.2$ & \multirow{2}{*}{0.5} & \multirow{2}{*}{0.1}\\
 & & post & 1925 & $2.03 \pm 0.03$ & $4.87 \pm 0.14$ & \\
\hline
\multirow{2}{*}{J0742-2822} & \multirow{2}{*}{55020} & pre & 253 & $2.1 \pm 0.4$ & $0.99 \pm 0.19$ & \multirow{2}{*}{0.4} & \multirow{2}{*}{0.4}\\
 & & post & 1708 & $2.01 \pm 0.09$ & $1.43 \pm 0.11$ & \\
\hline
\multirow{2}{*}{J0742-2822} & \multirow{2}{*}{56728} & pre & 1708 & $2.1 \pm 0.4$ & $1.43 \pm 0.11$ & \multirow{2}{*}{0.3} & \multirow{2}{*}{0.2}\\
 & & post & 1547 & $1.98 \pm 0.06$ & $1.44 \pm 0.07$ & \\
\hline
\end{tabular}}

\tablefoot{PSR is the pulsar name in J2000 epoch, GLEP is the glitch epoch expressed in MJD, $\Gamma_\text{S}$ is the photon index, $\Phi$ is the energy flux and $\Delta\Phi_{95\%}$ is the flux variation upper limit at 95\%.}
\end{table*}

\section{Conclusions} 
\label{sec:Spectral variability analysis}
Since its launch in 2008, \emph{Fermi}-LAT has detected almost 300 $\gamma$-ray pulsars, including dozens of young glitching pulsars. Our analysis focused on seven events: five glitches with $\Delta\Dot{\nu}/\Dot{\nu}>5.8\%$ (equivalent to or greater than the spin-down rate jump observed in J2021) and two glitches from the switching pulsar PSR J0742-2822. In our phase-averaged analysis, we find a consistency in flux and spectral parameters both before and after each selected glitch. Consequently, J2021 remains unique among the $\gamma$-ray pulsar population, as no other significant variability was observed during any of the glitching episodes investigated in this study. 

We computed relative upper limits on the flux variations spanning between $0.5\%$ and $40\%$. The upper limit of the PSR J1341-6220 glitch at MJD 55042 is compatible with the absolute energy flux. These limits constrain the magnitude of any possible $\gamma$-ray flux variations associated with glitches in these high spin-down-rate pulsars. The wide range of upper limits reflects the varying sensitivities of our analysis to different pulsars, which is influenced by factors such as the pulsar's brightness, the goodness of the spectral model, and the time span of observations. The lowest relative upper limit was found for the glitch of the Vela pulsar, indicating that this source is the most promising target for a dedicated search for variability. We stress that our work has thus far focused on a small subset of glitches. Moreover, no model is currently able to accurately predict the $\gamma$-ray pulsar state changes observed in J2021. A systematic search for $\gamma$-ray emission variations associated with glitches in pulsars would be promising and could ultimately lead to a deeper understanding of the physics of pulsar magnetospheres.

\begin{acknowledgements}

The \textit{Fermi} LAT Collaboration acknowledges generous ongoing support
from a number of agencies and institutes that have supported both the
development and the operation of the LAT as well as scientific data analysis.
These include the National Aeronautics and Space Administration and the
Department of Energy in the United States, the Commissariat \`a l'Energie Atomique
and the Centre National de la Recherche Scientifique / Institut National de Physique
Nucl\'eaire et de Physique des Particules in France, the Agenzia Spaziale Italiana
and the Istituto Nazionale di Fisica Nucleare in Italy, the Ministry of Education,
Culture, Sports, Science and Technology (MEXT), High Energy Accelerator Research
Organization (KEK) and Japan Aerospace Exploration Agency (JAXA) in Japan, and
the K.~A.~Wallenberg Foundation, the Swedish Research Council and the
Swedish National Space Board in Sweden.
 
Additional support for science analysis during the operations phase is gratefully
acknowledged from the Istituto Nazionale di Astrofisica in Italy and the Centre
National d'\'Etudes Spatiales in France. This work performed in part under DOE
Contract DE-AC02-76SF00515. This work was partially supported by Fermi GI grant 80NSSC23K1665.

This work is partially supported by the US National Science Foundation under Grant No. 2413060. Any opinions, findings, and conclusions or recommendations expressed in this material are those of the authors and do not necessarily reflect the views of the National Science Foundation.
\end{acknowledgements}

\end{document}